\documentclass[conference]{IEEEtran}
\IEEEoverridecommandlockouts
\usepackage{cite}
\usepackage{amsmath,amssymb,amsfonts}
\usepackage{algorithmic}
\usepackage{graphicx}
\usepackage{textcomp}
\usepackage{xcolor}

\usepackage{lipsum}

\newcommand\blfootnote[1]{%
  \begingroup
  \renewcommand\thefootnote{}\footnote{#1}%
  \addtocounter{footnote}{-1}%
  \endgroup
}

\def\BibTeX{{\rm B\kern-.05em{\sc i\kern-.025em b}\kern-.08em
    T\kern-.1667em\lower.7ex\hbox{E}\kern-.125emX}}
\begin{document}

\title{PatrIoT: IoT Automated Interoperability and Integration Testing Framework
\thanks{This research is conducted as a part of the project TACR TH02010296 Quality Assurance System for the Internet of Things Technology. The authors acknowledge the support of the OP VVV funded project CZ.02.1.01/0.0/0.0/16\_019 /0000765 “Research Center for Informatics”. Bestoun S. Ahmed has been supported by the Knowledge Foundation of Sweden (KKS) through the Synergi Project AIDA - A Holistic AI-driven Networking and Processing Framework for Industrial IoT (Rek:20200067).}
}

\author{\IEEEauthorblockN{Miroslav Bures}
\IEEEauthorblockA{\textit{Dept. of Computer Science, FEE} \\
\textit{Czech Technical University in Prague}\\
Prague, Czechia \\
miroslav.bures@fel.cvut.cz}
\and
\IEEEauthorblockN{Bestoun S. Ahmed}
\IEEEauthorblockA{\textit{Dept. of Math. \& Computer Science} \\
\textit{Karlstad University}\\
Karlstad, Sweden \\
bestoun@kau.se}
\and
\IEEEauthorblockN{Vaclav Rechtberger}
\IEEEauthorblockA{\textit{Dept. of Computer Science, FEE} \\
\textit{Czech Technical University in Prague}\\
Prague, Czechia \\
rechtva1@fel.cvut.cz}
\and
\IEEEauthorblockN{Matej Klima}
\IEEEauthorblockA{\textit{Dept. of Computer Science, FEE} \\
\textit{Czech Technical University in Prague}\\
Prague, Czechia \\
klimama7@fel.cvut.cz}
\and
\IEEEauthorblockN{Michal Trnka}
\IEEEauthorblockA{\textit{Dept. of Computer Science, FEE} \\
\textit{Czech Technical University in Prague}\\
Prague, Czechia \\
trnkamich@gmail.com}
\and
\IEEEauthorblockN{Miroslav Jaros}
\IEEEauthorblockA{\textit{Red Hat Czech s.r.o.} \\
Brno, Czechia \\
mjaros@redhat.com}
\and
\IEEEauthorblockN{Xavier Bellekens}
\IEEEauthorblockA{\textit{Dept. of Electronic and Electrical Engineering} \\
\textit{University of Strathclyde}\\
Glasgow, United Kingdom \\
xavier.bellekens@strath.ac.uk}
\and
\IEEEauthorblockN{Dani Almog}
\IEEEauthorblockA{\textit{Software Engineering Dept.} \\
\textit{Shamoon College of Engineering}\\
Be'er Sheva, Israel \\
almog.dani@gmail.com}
\and
\IEEEauthorblockN{Pavel Herout}
\IEEEauthorblockA{\textit{Dept. of CS and Engineering, FAS} \\
\textit{University of West Bohemia}\\
Plzen, Czechia \\
herout@kiv.zcu.cz}
}

\maketitle

\begin{abstract}
With the rapid growth of the contemporary Internet of Things (IoT) market, the established systems raise a number of concerns regarding the reliability and the potential presence of critical integration defects. In this paper, we present a PatrIoT framework that aims to provide flexible support to construct an effective IoT system testbed to implement automated interoperability and integration testing. The framework allows scaling from a pure physical testbed to a simulated environment using a number of predefined modules and elements to simulate an IoT device or part of the tested infrastructure. PatrIoT also contains a set of reference example testbeds and several sets of example automated tests for a smart street use case. 
\end{abstract}


\section{Introduction}

\color{blue}
Paper accepted at \textbf{IEEE International Conference on Software Testing, Verification and Validation 2021, Testing Tools Track}, virtual conference, April 12–16, 2021.
\newline
\newline
\textbf{https://icst2021.icmc.usp.br/track/icst-2021-Testing-Tool-Track}
\newline
\color{black}

The current dynamic development of IoT systems requires the development of proper quality assurance and testing methods to ensure a sustainable level of quality \cite{Marinissen2016}.\blfootnote{Bestoun S. Ahmed is also with Dept. of Computer Science, Czech Technical University in Prague, Czech Republic}
While a number of testing frameworks and testbeds have been developed recently (e.g., \cite{Datta_2018,Pontes2018,Sicari2017,Latre2016}), IoT testing is characterized by specific problems, which raise the demand for a universal and open solution to allow scalable interoperability and integration testing of IoT systems.

In this paper, we present PatrIoT\footnote{https://patriot-framework.io/}, an open automated testing framework that allows a flexible composition of testbeds from both simulated and physical devices in addition to parts of the infrastructure. The framework is developed in a joint academia-industry collaboration project between the Czech Technical University (CTU) in Prague and Red Hat, as the industry partner, with external consultancy cooperation of the University of West Bohemia, University of Strathclyde, UK, and Shamoon College of Engineering, Israel. The PatrIoT framework is accompanied by a set of example testbeds and automated tests created for modeled smart street IoT applications, which eases its applicability for industry engineers in various IoT projects.

\section{Related Work}\label{section:relatedwork}

Regarding the construction of IoT automated testing frameworks, the greatest effort is currently dedicated to the security and privacy aspects~\cite{Kiruthika2015}. Several testing frameworks have been developed recently in these areas, e.g.,~\cite{Datta_2018,Pontes2018}. In addition, other studies focused on domain-specific frameworks for industrial IoT systems~\cite{SeokcheolLee2017}, healthcare systems~\cite{Sicari2017}, or smart cities~\cite{Latre2016,Lanza2015}. Additionally, Kim and Ziegler~\cite{KimEE2017} addressed the challenges of IoT testing in the area of interoperability and conformance. For several industrial projects today, extensive testing strategies of devices and their interoperability and interactions are required. The magnitude of the problem increases with the number of various devices that communicate with each other, the versions of their firmware, and the various communication protocols used. This variety creates an extensive number of combinations that lead to a combinatorial explosion in the number of test cases. 

Regarding the construction of testbeds, a significant number of prototypes have been reported. The prototypes mainly depended on the IoT domain and the purpose of testing. They vary from general testbeds \cite{JunHwanHuh2017,Lanza2015} to specific testing strategies such as protocol and scalability testing \cite{Rodrigues2017} or system security testing \cite{Choi2016}.

Test automation frameworks and testbeds, specifically focusing on interoperability and integration testing, are discussed in several studies. Wu~\textit{et al.} summarized currently available approaches to test Wireless Sensor Networks (WSNs) and include integration testing in their discussion~\cite{Wu2019}. However, the study is not exhaustive, as other possible approaches can be identified and employed.

Pontes~\textit{et al.} have recently presented a pattern-based test automation framework for integration testing of IoT systems. In their proposal, interoperability issues such as various communication protocols are tackled~\cite{Pontes2018}, but overall, the concept focuses rather on physical-devices-based testbed rather than on a mixture of physical devices and their simulations.

Another study by Rosenkranz~\textit{et al.} discusses issues arising from the heterogeneity of IoT systems and their interoperability testing and proposes a high-level concept of interoperability testing framework~\cite{Rosenkranz_2015}. The concept focuses on the integration of standalone IoT systems and principally, simulation of individual components is not emphasized.  

Considering the analyzed studies, we can conclude that the general support for IoT interoperability and integration testing that allows flexible scaling from simulations to physical testbed configuration is not discussed sufficiently in the literature.

\section{PatrIoT Framework}
\label{section:framework}
In our explanation of the PatrIoT framework, we first summarize its design principles. We follow with a description of its structure, enabling efficient and scalable automated interoperability and integration testing for IoT solutions.

\subsection{Design Principles}
\label{section:design_principles}

During the development of the PatrIoT framework, we considered the following design and conceptual principles to increase the applicability and flexibility of the framework.

\begin{enumerate}
\item The framework is developed as a flexible and open solution under a nonrestrictive open-source license Apache 2.0.

\item The framework employs current established open-source resources as much as possible. The goal of using these defacto standards is to ensure a better learning curve, better applicability, and better integration potential with other frameworks or parts of the testing infrastructure.

\item The framework is developed to be scalable to support a wide range of interoperability and integration testing, such as application programming interface (API) testing and complex end-to-end (E2E) integration testing.

\item The framework allows a flexible composition of physical and simulated items to deal with and substitute for the nonexistence of some IoT items that are not ready yet due to hardware cost, for example.

\item Simulation of the devices is solved as a modular system in which a particular device is used in the testing framework as a configurable module. Predefined sample devices are provided to the framework users to be used as templates.

\item The framework allows the connection of data inputs from Model Based Testing (MBT) tools via open interfaces. Namely, these interfaces are provided for combinations of test data inputs \cite{Ahmed2017} and path-based test cases \cite{bures2019employment}.

\item The framework supports the creation of well-structured and organized automated interoperability and integration tests, which minimize the maintenance overhead that may arise by keeping the test cases up-to-date when a System Under Test (SUT) is being changed. Similarly as in web applications \cite{bures2015metrics} for instance, the IoT would is not an exception in this point.

\item The framework is constructed as a set of cooperating reusable modules that are connected via open interfaces. This gives the individual parts of the framework the possibility to be integrated with the existing parts of its users’ testware.

\end{enumerate}

\subsection{The Main Components of the PatrIoT Framework}\label{section:main_components}

The framework comprises several principal components. The responsibilities of these components are as follows:

\begin{itemize}
\item The \textbf{test runner} represents the main user interface so that the users may control the testing process in the framework. This component executes the defined automated test cases in a configured testbed and ensures the running of all auxiliary activities related to this execution. The test runner also unifies access to different parts of the testbed (e.g., network simulation manager or device emulator) into one scripting environment. The test runner is composed of two main subcomponents:

\begin{itemize}
\item The \textbf{integration test runner} is responsible for the execution of the defined automated integration test suites.

\item The \textbf{performance test runner} is responsible for conducting a performance testing of the SUT. This process also involves metrics collection to support the evaluation of performance testing.
\end{itemize}

\item The \textbf{testbed hub} is responsible for the orchestration of the physical and simulated devices' behavior with the infrastructural parts of the testbed. The hub transmits events and actions to the individual devices and elements of the testbed environment. For example, the change in the network topology, the evolution of the data generated by a simulated component, or the disconnection of the component from the network. The hub also ensures the monitoring of the test environment.

\item The \textbf{network simulation manager} is responsible for the emulation of the network connections between the components. The component provides interfaces that allow the definition and deployment of selected network topologies and their management.

\item The \textbf{device emulator} provides the capability for emulating various types of physical IoT devices as well as the generation of different kinds of data by probabilistic algorithms. This allows the emulation of the realistic behavior of such devices. We present more details for the simulated supported devices in Section \ref{section:simulation_of_devices}.

\item The \textbf{collector} is responsible for data collection from the testbed environment (namely, network, device emulators and other components). The collector records details about these data to allow the rerunning of the test cases in the same conditions and scenarios to reproduce the detected defects. This data recording process is essential because the simulated IoT devices may produce a random torrent of data.

\item The \textbf{provisioner} is responsible for defining, deploying, setting up, and initializing the testbed. The provisioner allows storing multiple configurations of the testbed and initializing selected configurations.

\item The \textbf{reporter} is responsible for processing and presenting the test results and the events that occurred within the testbed during the testing process. This component aims to provide a unified reporting interface, which accepts the test results and events and states from the testbed to maintain the context information needed for effective defect analysis.

\end{itemize}

\subsection{The Network Infrastructure Simulation of the SUT}\label{section:network_simulation}

As mentioned previously, the PartIoT framework testbed can be physical parts of the network infrastructure and/or combined with a simulated network infrastructure. The components' configurations of this simulation are performed in the Docker environment. Here, the configuration of the physical IoT devices in the testbed is connected to the physical part of the network infrastructure. The virtual parts of the testbed (i.e., the simulated IoT devices, the simulated active elements of the network as gateways, and other infrastructural elements) are deployed as individual Docker containers. The individual links of the simulated network are configured as virtual links between the Docker containers or the Docker container and the interface to the physical part of the network. Within this system, a flexible scaling of the testbed configuration is achieved, which leads to an easy deployment mechanism by the Docker infrastructure.

\subsection{The Simulation of IoT Devices}\label{section:simulation_of_devices}

The framework provides a set of reusable building blocks from which a particular IoT device simulation can be assembled. In its current version, the framework provides two principal types of simulation, the data provider and the actuator.

The \textbf{data provider} can be assembled from a set of predefined building blocks in three levels: data generator, data transformer, and connector. The \textit{data generator} produces a stream of data, which simulates the data produced by a physical data provider (e.g., a sensor connected to the network). A particular flow of data is driven by a configuration based on a defined probabilistic function, for example, the data for the temperature in time or the GPS trajectory in time. Subsequently, the \textit{data transformer} converts the stream of data produced by the data generator to a format in which the device provides the data to the IoT system (e.g., JSON format). Finally, the \textit{connector} connects the simulated device to the rest of the IoT system and ensures the transmission of the data.

The \textbf{actuator} simulates a more complex IoT device, which reacts to the API calls from the rest of the system and returns the result or the feedback data after performing a particular action triggered by the API call. This simulated device can be assembled from a set of predefined building blocks in three levels: interface, internal state machine, and connector. The \textit{interface} defines the API of the simulated device in a format in which the device communicates with the rest of the IoT system (e.g., JSON or SOAP). The \textit{internal state machine} defines a simplified behavior of the simulated component, which reacts to the device API calls and responds by a particular data that is provided as the output of the particular API. The \textit{connector} connects the simulated device to the rest of the IoT system and ensures the ability to connect to the API of the simulated component.

In addition to the building blocks in the above categories, the framework contains a set of sample predefined simulated devices. These samples include several simulated data providers as motion detection sensors, light level sensors, CO and CO$_2$ levels sensors, temperature sensors, and fire outbreak detectors. Additionally, several examples of simulated actuators for smart-home appliances are available.

\subsection{Implementation Details}

In the implementation of the framework, we followed the rule to reuse and utilize the already established open-source components. We used a modified version of the JUnit 5\footnote{https implementation integration test runner://junit.org/junit5/} framework. Since the current version of the JUnit framework is insufficient for our requirements, we added several extra features to the framework to specifically support automated interoperability and integration testing. These features involve additional support for the synchronization and orchestration, such as the implementation of warning state in the automated test, interruption possibility in a situation where the test flow is broken, and more flexible orchestration possibilities of the test step flow.

We implement the performance test runner using the already established open-source community project Perfcake\footnote{http://perfcake.org/} to integrate and adapt with the PatrIoT framework for IoT testing specifics. The network simulation manager is implemented as a component to control the Docker\footnote{https://github.com/docker} container deployment process and set up the respective network interfaces for the deployed containers. The containers are controlled by a rich API written in Java that allows the user to specify in the code how the software should be built and deployed. It also provides an API for the network topology definition, which spawns several virtual networks interconnected by software routers and connects the SUT to appropriate places. Device emulators and all other components of the framework are implemented as open-source Java 8 classes in a modular structure that allows future flexible extensions of the framework and its integration with third-party testing tools and environments. The collector component provides an open interface to Elasticsearch\footnote{https://www.elastic.co/products/elasticsearch} solution to allow further processing, analysis, and visualization of the test results and related context information from the testbed. The reporter component is created by an adaptation of the current Maven Surefire plugin\footnote{https://maven.apache.org/surefire/maven-surefire-plugin/}, which is currently used for test reporting in the JUnit platform.

\section{Industrial application}
\label{section:results}

The PatrIoT framework is currently being employed in several IoT projects, which confirm the applicability of the proposed concept and technical solution. In this section, we summarize (1) the general feedback from these projects, (2) results from the application of the PatrIoT in the IoT middleware communication infrastructure Active Messaging Queue (AMQ Online), a product of Red Hat, which is employed in a number of current solutions, for instance by Bosch, Swiss Federal Railways (SBB), British Petrol or Intermountain Healthcare, and, (3) we describe a demo project of a smart street system, which documents capabilities of the framework and serves its users to learn the framework and apply it.

\subsection{General Feedback}

The general feedback from the pilot applications confirms the very good flexibility of the framework in the composition of a mixed testbed, which is composed of physical and simulated IoT devices, as well as very good scalability between these two options. The fact that the framework is built using JUnit as its basic test runner component also creates several benefits, as the pilot projects show. As JUnit is known as the defacto standard for unit testing and close-API integration tests, the framework is applicable to a wide spectrum of automated tests, including pure unit tests, various types of integration tests, and user-interface-based automated tests using the JUnit framework as its runner.

As examined during the projects, the heterogeneity of the programming languages (Java as the core framework language) versus other languages, e.g., Python, in the case of rescue mission planning and management systems, limits the application of the PatrIoT for the classic unit tests of the SUT; however, for integration testing and user-interface-based automated tests, the capabilities are not limited. In such cases, individual SUT devices and modules act as black boxes, exposing APIs that are exercised by the automated tests run by the PatrIoT framework.

The necessity to understand the tested API and connect to a tested device or module by an available interface represents, based on the experience from the case studies, a crucial key point, allowing effective testability of the individual system parts. However, this is a general issue in IoT test automation that cannot be influenced by the test automation style or concept.

Limits of the framework might be observed in the case of event synchronization of the physical and simulated devices. The framework provides options for synchronization of events in the IoT system during the test via test runner and testbed hub components; however, for the more complex and precise synchronization of events, the developer of the automated tests needs to handle the behavior explicitly in the automated test code. More extensive support for the complex synchronization cases is included in the development roadmap of the framework.

\subsection{IoT Middleware Communication Infrastructure}
\label{subsec:amq}

AMQ Online is Red Hat product for managed, self-service messaging on Kubernetes and OpenShift. AMQ Online can run on user-provided infrastructure or in the third party provided cloud solutions and simplifies running a messaging infrastructure for IoT solutions.
AMQ Online can be used for many purposes, such as moving an IoT messaging infrastructure to the cloud without depending on a specific cloud provider or building a scalable messaging backbone for an IoT system. The solution is employed by several end users in their solutions, for instance, Bosch, Swiss Federal Railways (SBB), British Petrol and Intermountain Healthcare.

During the fourteen months of the pilot project of PatrIoT application in the AMQ Online system, the framework was adopted by AMQ development and testing team and included in its test automation infrastructure, serving as a complement to established unit tests. Automated integration tests for AMQ Online use cases were prepared, including close-API integration tests as well as complex integration scenarios.

The initial effort of AMQ development and testing team needed to get familiar with the framework was 6 Man-days (MDs), including preparation of the test automation plan for new integration tests.

We evaluated two latest releases of the AMQ Online product: release 1.4.0 (denoted as REL1 further) and release 1.5.0 (denoted as REL2 further).

Over the pilot project run, compared to previous manual execution of integration tests, requirements coverage increased by 35\% for REL1 and further increased by 24\% for REL2. Increase in line and branch coverage of the code was slightly increased; however, due to the previous high level of code line coverage, which was 92\%, additional execution of some code lines by new integration tests was not relevant to evaluate.

The time needed to prepare PatrIoT-based automated integration tests was 14 MDs in REL1 and 3 MDs in REL2. In REL 1, 25 complex integration test scenarios were created, consisting of 173 test steps. For REL 2, additional seven complex integration test scenarios were created, consisting of 56 test steps.

In both of individual releases, three builds with automated regression tests were conducted. The effort needed for one round of automated tests using the PatrIoT framework was 2 MDs. Most of this effort spent on failure analysis and defect reporting. In contrast, the effort needed to one round of regression tests conducted by the previous manual way of integration testing was 8 MDs. 

The collected data verify the economic viability of test automation using the PatrIoT framework - Test automation in one release gained 36 MDs savings, which overweight the total 17 MDs needed for the preparation of the tests.

Regarding the effectiveness of the PatrIoT-based automated tests, compared to the 18 total regression defects detected by unit tests and manual integration testing, eight new defects were detected by PatrIoT in REL1 and 13 in REL2. By the severity of these defects for both releases, they included two blockers, eleven critical defects, one major defects, and six minor defects.

Considering the actual state of AMQ Online, which is well covered by unit tests and regular regression tests discover tens of defects maximally, the application of PatrIoT automated integration tests was effective in finding new defects present in the system.

\subsection{Demo Smart Street Testbed}

Discussed features of the framework are also documented in another project, the demo smart street testbed, resulting in a set of open examples that are available for the framework's users. The examples further demonstrate the capability of the framework to flexibly scale from a purely physical testbed to a completely simulated IoT infrastructure. The most important case is when the testbed is mixed from physical and simulated devices.

\begin{figure}
\centering
\includegraphics[width=.48\textwidth]{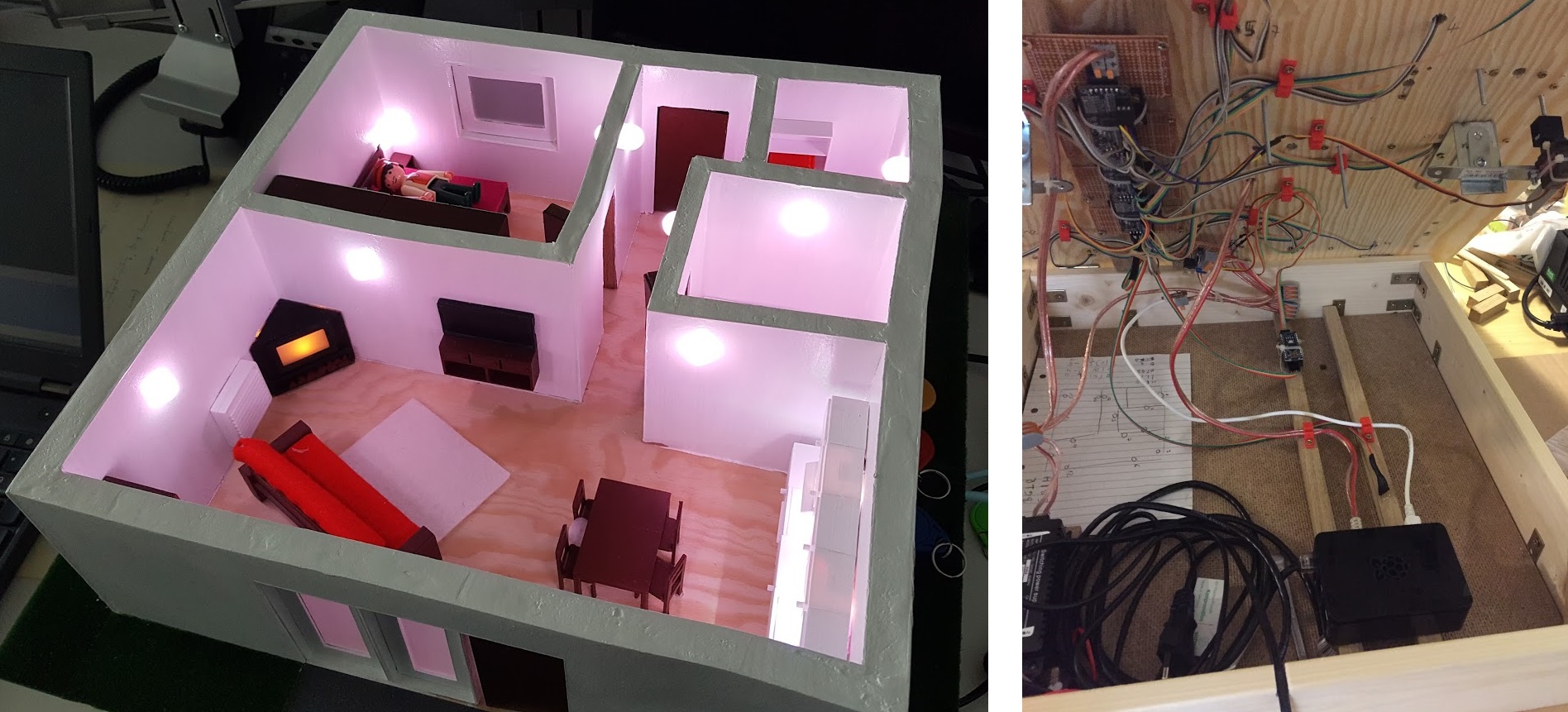}
\caption{A prototype of the physical smart house model}
\label{figure:house_model_final}
\end{figure}

The smart street model is composed of a set of smart houses (each represented by a set of sensors and actuators), gateways of the smart houses, and a controlling system for the smart street. They are implemented as a server service. The physical device of the smart house model contains actuators for central heating, air conditioning, lights, home sound system, opening the garage doors, and automated opening of windows. Moreover, it contains sensors for room temperature, air humidity, and an RFID tracking system. The simulated device of the smart house, in addition to these devices, comprises more actuators, such as automatic curtains on the windows, opening the house gate, and home appliances (e.g., a coffee machine) connected to the network. Additionally, the set of sensors is enriched by other devices for motion detection, light level, CO and CO$_2$ levels, outside temperature, and fire outbreak detectors. To make the testbed example available in the framework and independent of a real smart home deployment, a physical model of the smart house is available (its prototype is presented in Figure \ref{figure:house_model_final}).

Table \ref{table:configurations} presents a set of available sample testbed configurations within the framework. We presented the possible scalability of the framework to support testbeds from a purely physical configuration to purely simulated testing environments.

\begin{table}[htbp]
\caption{Sample testbed configurations available in the PatrIoT framework}
\begin{center}
\begin{tabular}{|c|p{5.3cm}|p{1.9cm}|}
\hline
ID & Configuration  & Testbed Type\\
\hline
\#1 & One smart house, all devices physical, physical gateway & Purely physical testbed \\
\hline
\#2 & One smart house, all devices physical, simulated gateway & Mixed testbed \\
\hline
\#3 & Two smart houses and smart street server. House A: all devices physical, physical gateway. House B: all devices simulated, simulated gateway, smart street system simulated     & Mixed testbed \\
\hline
\#4 & Two smart houses and smart street server, all devices and gateways of House A and House B simulated, smart street system simulated & Purely simulated testbed \\
\hline
\end{tabular}
\label{table:configurations}
\end{center}
\end{table}

The testbeds are accompanied by a set of automated sample test cases that evaluate a set of defined use cases of the smart street. The templates also demonstrate the recommended reference architecture for creation of automated tests.

\section{Discussion}\label{section:discussion}

The concept of universal and flexible test automation and testbed construction framework is promising. There are two issues to be discussed in this context. Conceptually, a question can be raised if considerable flexibility in mixing the physical devices testbed with the simulated devices is needed, and if the problem can be solved by a simple strict application of the V-model. In other words, the individual parts can be tested at the unit level, integration interfaces at the unit integration level, and then the units are combined, and the final configuration can be tested end-to-end using a lighter testbed configuration. This approach sounds like a logical option, and several studies reported this approach (e.g., \cite{Kiruthika2015,Marinissen2016,Xu2014}). Our industry project observations from interviews with 15+ IoT solution providers suggest that such an approach can reach its limits. The premise “if units work and individual interfaces also work, then the E2E process shall also work” is not simply valid, as many practitioners have documented numerous examples of actual tricky interoperability and integration defects.

Another concern may be raised regarding the construction of a universal test automation framework for IoT systems. Considering the specifications of various systems, various protocols are currently employed, and the nature of the individual systems and the variety of user devices may make this goal seem too ambitious. However, this concern can be minimized by the open architecture of the testing framework to allow the connection of various devices and infrastructural parts by adding proper communication adapters between the test cases and the devices. Additionally, adding new types of simulated devices, which are configurable from predefined building blocks, will minimize that concern. Such flexibility is the goal of the PatrIoT framework; however, to be objective, it may reach its limits for some IoT systems or their specific parts.

\section{Conclusion}\label{conclusion}

The increasing demand for effective interoperability and integration testing of IoT solutions and the issues related to the testbed design and construction also implies a demand for open and flexible test automation frameworks to support this process. The construction of a universal test automation framework is a significant industrial challenge. The related problems start with the composition of the testbed. As it might be unrealistic to create a testbed that is composed of solely physical devices, the simulation of the devices arises as a prospective option. The construction of these frameworks and simulators can naturally tend to a proprietary solution with low reusability of the framework. Another possible issue is low-level flexibility to enable the combination of physical and simulated devices. All these issues are addressed by the PatrIoT framework, which we presented in this paper. The framework allows the creation of a flexible environment to construct a testbed dynamically from a mixture of physical devices and infrastructure parts and particular device simulation. The simulators can be configured using predefined modules of the framework. In this process, a set of predefined examples can be used. 

The pilot application of the framework showed the economic effectiveness of the created automated tests in terms of detected defects as well as the economic viability of the test automation process. As part of our future work, we are extending the framework to support even more complex cases of synchronization of events during the tests. We are also adding more modules that allow the creation of simulators of various IoT devices. The next prospective direction is adding more capability of the framework for agile testing means, as we successfully tried for web-based systems \cite{bures2018tapir}.

\bibliographystyle{IEEEtran}
\bibliography{references}

\end{document}